\documentclass[conference]{IEEEtran}
\usepackage{comment}
\usepackage{multirow}
\usepackage{siunitx}
\usepackage{url}
\usepackage{color}
\usepackage{setspace}
\usepackage{rotating}
\usepackage{graphicx}
\usepackage{caption}
\usepackage{subcaption}
\usepackage{tabularx}
\usepackage{booktabs}

\graphicspath{{figures/}} 

\newcommand{\cellname}{Threshold-Dependant Camouflaged Cells}
\newcommand{\abb}{TD camouflaged cells}
\newcommand{\abbsingular}{TD camouflaged cell}

\begin{document}
%


\title{Threshold-Dependent Camouflaged Cells to Secure Circuits Against Reverse Engineering Attacks\vspace{-0.3in}}


\author{\IEEEauthorblockN{Maria I. Mera Collantes, Mohamed El Massad, Siddharth Garg}
\IEEEauthorblockA{Department of Electrical and Computer Engineering\\
New York University\\
mimera@nyu.edu, me1361@nyu.edu, sg175@nyu.edu}
}


%


\maketitle

\begin{abstract}

With current tools and technology, someone who has physical access to a chip can extract the detailed layout of the integrated circuit (IC). By using advanced visual imaging techniques, reverse engineering can reveal details that are meant to be kept secret, such as a secure protocol or novel implementation that offers a competitive advantage. A promising solution to defend against reverse engineering attacks is IC camouflaging. In this work, we propose a new camouflaging technique based on the threshold voltage of the transistors. We refer to these cells as threshold dependent camouflaged cells. Our work differs from current commercial solutions in that the latter use \textit{look-alike} cells, with the assumption that it is difficult for the reverse engineer to identify the cell's functionality. Yet, if a structural distinction between cells exists, then these are still vulnerable, especially as reverse engineers use more advanced and precise techniques. On the other hand, the proposed threshold dependent standard cells are structurally identical regardless of the cells' functionality. Detailed circuit simulations of our proposed threshold dependent camouflaged cells demonstrate that they can be used to cost-effectively and robustly camouflage large netlists. Corner analysis of process, temperature, and supply voltage (PVT) variations show that our cells operate as expected over all PVT corners simulated.

\end{abstract}


%
\IEEEpeerreviewmaketitle

\section{Introduction}

The need to protect the intellectual property (IP) of integrated circuit (IC) design companies is increasing as threats from IC reverse engineering attacks become more prominent.
Reverse engineering of ICs consists of using advanced depackaging, delayering and imaging techniques in order to acquire the gate-level netlist. Not only is it possible to successfully obtain the full layout and netlist of complex ICs fabricated using advanced processes~\cite{intelxeon}, it is also commercially feasible, such that companies such as Chipworks~\cite{chipworks} and Degate~\cite{degate} offer these services. IC reverse engineering can cause the competitive advantage of a chip maker to diminish, as well as reveal details about the chip that are meant to be secret, for example, the implementation of a secure communication protocol~\cite{applecable}.

IC camouflaging technology has been proposed~\cite{syphermedia,RSS+13} as a way to counteract reverse engineering attacks. With this countermeasure, the designer aims to hinder the ability of the reverse engineer to visually decipher the functionality of a gate. For example, NAND and XOR gates are made visually similar by fabricating them with identical metal and polysilicon layers and contact locations.  
The difference lies in what type of contacts are placed. These contacts may either be \textit{true} or \textit{dummy}. The former electrically  connects the two layers, and the latter has a small gap in the middle so that it behaves like an open circuit. The selection of which type of contact to use establishes the functionality of a gate.

The principal assumption in this method is that, due to imprecision in the etching process~\cite{syphermedia,RSS+13}, the attacker is unable to discriminate between true and dummy contacts. Consequently, the Boolean functionality of the camouflaged gate cannot be ascertained by exclusively using traditional reverse engineering techniques. This method is referred to as \textit{dummy contact based camouflaging}.

The expectation that a reverse engineer's capabilities and techniques will not improve and become more sophisticated and precise reveals a weakness in dummy contact based camouflaging. Indeed, ICs fabricated in advanced technology nodes down to 22 $nm$ have been reverse engineered~\cite{intelxeon}. This illustrates that while visually discernible differences in layout structure exist, the threat from IC imaging based reverse engineering attacks will remain.  

In this paper, we propose a \emph{new} camouflaging 
technique that addresses this vulnerability. 
We present an approach that leverages the intrinsic characteristics of the material used instead of the physical structure of a camouflaged gate. For this reason the gate layout, including vias and contacts, is the same regardless of its Boolean functionality. Instead, \emph{the functionality of the proposed camouflaged gate is determined by the {threshold voltage} of its transistors}, that is, depending on whether it has been fabricated using a high-$V_{th}$ or low-$V_{th}$ process. As such, this technique uses the standard features available in commercial CMOS dual-Vth (or multi-Vth) processes~\cite{kao1997transistor}. We call this technique \textit{threshold dependent (TD) camouflaging} and the cells are called TD camouflaged cells. To the best of our knowledge, our work is the first to present a camouflaging solution that leverages commercial multi-Vth CMOS processes to camouflage the Boolean functionality of a gate depending on whether it has been fabricated using high-$V_{th}$ or low-$V_{th}$ transistors.  

There are several advantages that result from using \abb{}. First, traditional etching and optical imaging techniques used by reverse engineers are ineffective since the functionality of the cell depends on the threshold voltage, which in turn is determined by the doping concentration of the cell. Second, even though an attacker can image and identify the doping concentration of a cell using methods such as spreading resistance profiling (SRP) and secondary ion mass spectrometry (SIMS), these have limited spatial resolution~\cite{huang1996direct}. Other approaches involve using scanning capacitance probes. While these can offer higher spatial resolution they still have limited accuracy because they do not directly measure dopant concentration~\cite{huang1996direct}. 
As Chipworks notes, physical features can potentially be reverse engineered deep into the nanometer scale ``...with high-resolution electron microscopy, but obtaining details of the chemical composition of the structure is now in the realm of counting atoms"~\cite{torrance2007reverse}. 
 
Our approach can be used in conjunction with a multi-$V_{th}$ process~\cite{kao1997transistor} in a cost effective manner. That is, each cell can be manufactured using either high-$V_{th}$ or low-$V_{th}$, depending on the gate that it is replacing, 
which means that camouflaging $k$ cells results in up to $2^k$ different possible functions. 
Lastly, we note that \abb{} can be used in \emph{conjunction} with dummy contact based camouflaged 
technique to make the reverse engineering problem even harder for an attacker.

\section{Contributions}
\label{sec:contribution}

The novel contributions of this work are:
\begin{itemize}
\item Proposal of \abb{}, a logic cell whose Boolean functionality depends on its threshold voltage,

\item design, analysis and evaluation of a specific instance of a \abbsingular{} that functions as either an AND or an OR gate depending 
on its threshold voltage, and

\item experimental results showing the robustness of \abb{} 
over process, voltage and temperature (PVT) corners, and 
comparing the delay and power overheads of \abb{} to conventional 
logic families.

\end{itemize}

The rest of this paper is organized as follows. 
In Section~\ref{sec:rel}, we describe previous work related to our paper. Section~\ref{sec:method}, presents and discusses the design of \abb{} as well as considerations for robustness to process variations. We report experimental results in Section~\ref{sec:exp}. Finally, we conclude our paper in Section~\ref{sec:conc}.

\section{Related Work}\label{sec:rel}

The idea of controlling the functionality of a cell by using varying operating parameters has been previously proposed. 
For instance,~\cite{sekanina2009repomo32,stoica2001polymorphic,ruzicka,poly} explore the use of \emph{polymorphic gates}. 
These cells are characterized by having different functionalities that change according to the variation of supply voltage or temperature. Since controlling temperature is complicated and determining the supply voltage of each gate is feasible, these cells cannot be used for IC camouflaging. 

A previous notion that has been explored in the context of IC camouflaging is the use of look-alike cells~\cite{syphermedia,RSS+13}.
The functionality of these cells depend on the location of true and dummy contacts which are claimed to hinder the capability of the attacker to identify which vias are functional. Even though this technique is used to prevent the reverse engineering of an IC, it still relies on a physical structural difference, which may not deter a more skilled attacker.

Our proposed \abb{} are structurally identical, yet their function depends on an inherent characteristic of the transistor. Hence the attacker would require more sophisticated tools to ``image" the threshold voltage of each cell since this is dependant on channel doping. Our cells can be used in addition to or instead of conventional dummy-contact based camouflaged cells.

There is some prior work 
involving threshold voltage based camouflaging~\cite{larson1992convertible,walden1993dynamic}. 
The first technique~\cite{larson1992convertible} uses selective ion implantation in certain 
transistors to ensure that these transistors become permanently
``stuck-at" either logic 0 or logic 1 (essentially converting transistors 
to \emph{permanently} on/off switches). This introduces two 
problems: (1) the required increase in 
channel doping is large, making it easier for an attacker to detect;
and (2) 
the approach requires extra steps in the current CMOS flow. 
In contrast, \abb{} leverages existing, commercial 
multi-Vth processes, and requires much subtler changes in 
doping density. The latter~\cite{walden1993dynamic} technique uses post-fabrication focused-ion beams (FIB) to alter doping concentration. This must be 
done individually for each chip, making it a slow and expensive process. Concurrently with our work, 
another approach for threshold-dependent camouflaging has been 
recently
proposed. However, this approach 
requires the generation and distribution of 
additional precise voltages on the chip (besides VDD and ground),  
while having similar overheads as our technique~\cite{Iyengar15}. 
Since the paper is currently an Arxiv draft made available only very recently, a thorough empirical comparison is left as future work.

Finally, note that recent work has also proposed 
stealthily changing the polarity of the diffusion layer 
for both attack~\cite{becker} and defense~\cite{dpd}. 
In this, 
the \emph{polarity} of the diffusion layer is changed. 
Attacks on these techniques that are able to 
reverse engineer the polarity of a diffusion layer (p- or n-type)~\cite{dpd-attack} would not work for \abb{} since we only change the doping concentration, not polarity.

Given access to a camouflaging technique, prior work has also addressed the problem 
of which gates to camouflage so as to make it difficult for the attacker to reverse 
engineer their identities given the input/output behaviour of the chip~\cite{RSS+13}. 
A recent attack on this technique has demonstrated that the number of gates that needs 
to be camouflaged is much larger than previously thought~\cite{ndss2015}. 
Regardless, the question of which and how many gates in the netlist to camouflage 
is orthogonal to the work presented in this paper. 
\color{black}

\section{\abb{}}\label{sec:method}

\subsection{Basic Operating Principle}

\begin{figure}[htpb]
\centering
\includegraphics[width=3in, height=1.3in]{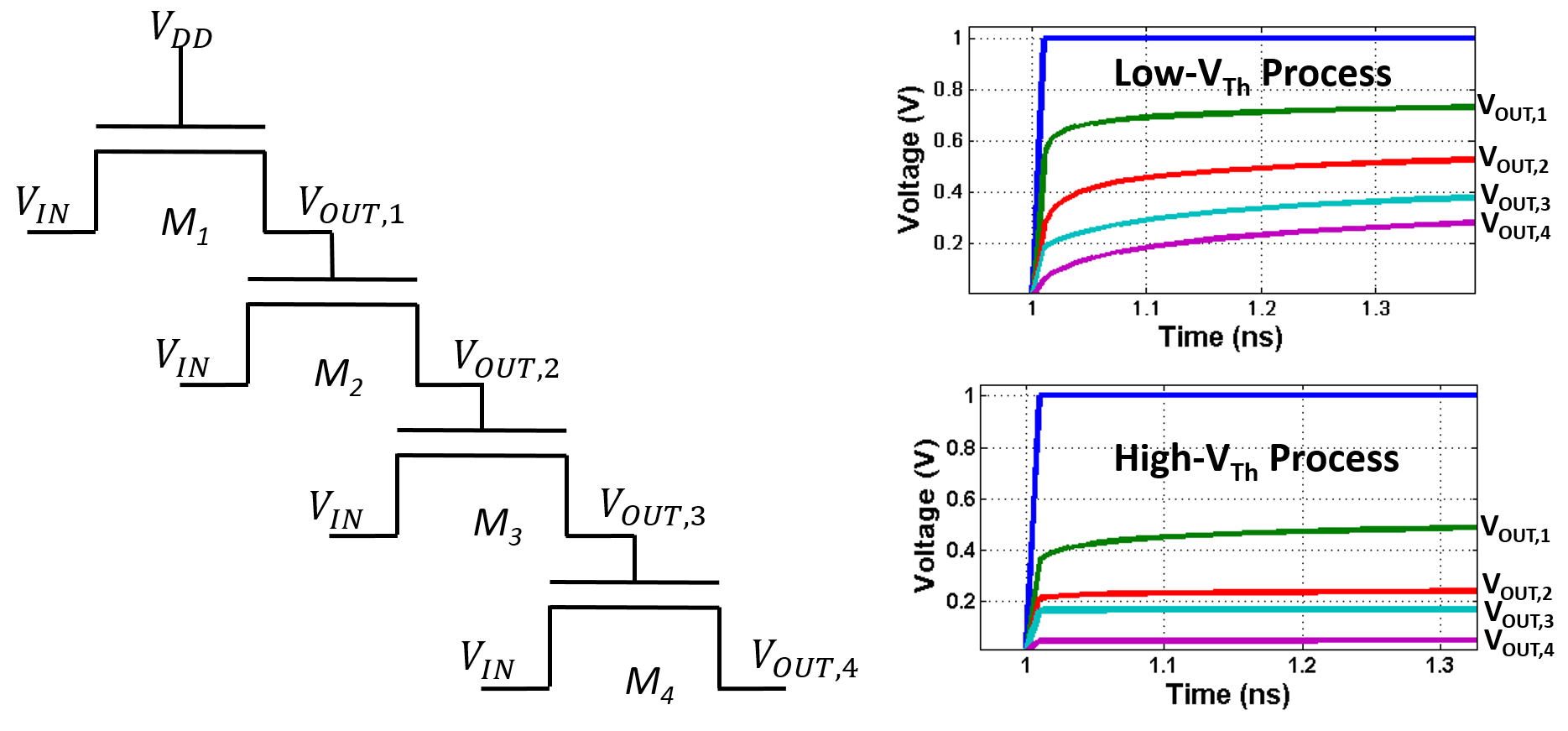}

\caption{(Left) A cascade of four NMOS level-converting pass 
transistors. (Right) Output of each stage with transistors fabricated in Low-$V_{Th}$ versus High-$V_{Th}$ process. Observe that the output 
voltage levels for the latter are lower than the corresponding voltage levels for the former. 
}

\label{fig:cascade-concept}
\end{figure}

When transistor pass gates are connected in a cascade layout like in Figure~\ref{fig:cascade-concept} (Left), 

the voltage at $V_{OUT}$ drops after each stage. 
In this layout, if $V_{IN}$ is set to $V_{DD}$, then, the maximum voltage at $V_{OUT,1}$ will be $V_{DD} - V_{Th}$. Consequently, if $V_{IN}$ of $M_2$ is set to $V_{DD}$ then, $V_{OUT,2}$ will be $V_{DD} - 2~\times V_{Th}$. As we continue to connect transistors in this way we can see that the voltage at the source can be characterized by $V_{DD} - n~\times V_{Th}$, where $n$ is the number of stages of connected transistors. We make these assumptions for an ideal NMOS device with no sub-threshold leakage. Since $V_{Th}$ depends on the doping concentration of the transistor, the voltage at the source of $M_n$ varies according to the threshold voltage asserted during the fabrication process.

Figure~\ref{fig:cascade-concept} (Left) shows the layout in which the $n=4$ NMOS transistors are connected as well as the output voltage at each source terminal for for each stage obtained from a simulation using Predictive Technology Models (PTM)~\cite{ptm} for a 22 $nm$ technology for High-$V_{Th}$ and for Low-$V_{Th}$ processes. For each simulation all transistors in the layout were considered to have either High-$V_{Th}$ or Low-$V_{Th}$. In Figure~\ref{fig:cascade-concept} (Right) we can see that the output voltage decreases at each stage, and more so for High-$V_{Th}$ simulations than for Low-$V_{Th}$. Note that the difference between the theoretically expected and observed simulated results is due to the assumption of ideal transistors with  

zero off-current, when, for 22 $nm$ technology, sub-threshold conduction is non-negligible.

\subsection{\cellname{} - AND/OR}
\subsubsection{Schematic Design}

Using the previous principle we have developed a layout of a gate that can have the functionality of AND or OR depending on the threshold voltage that is asserted during the manufacturing process. \vspace{-2ex}

\begin{figure}[htpb]
\centering
\includegraphics[width=2.7in, height=1.8in]{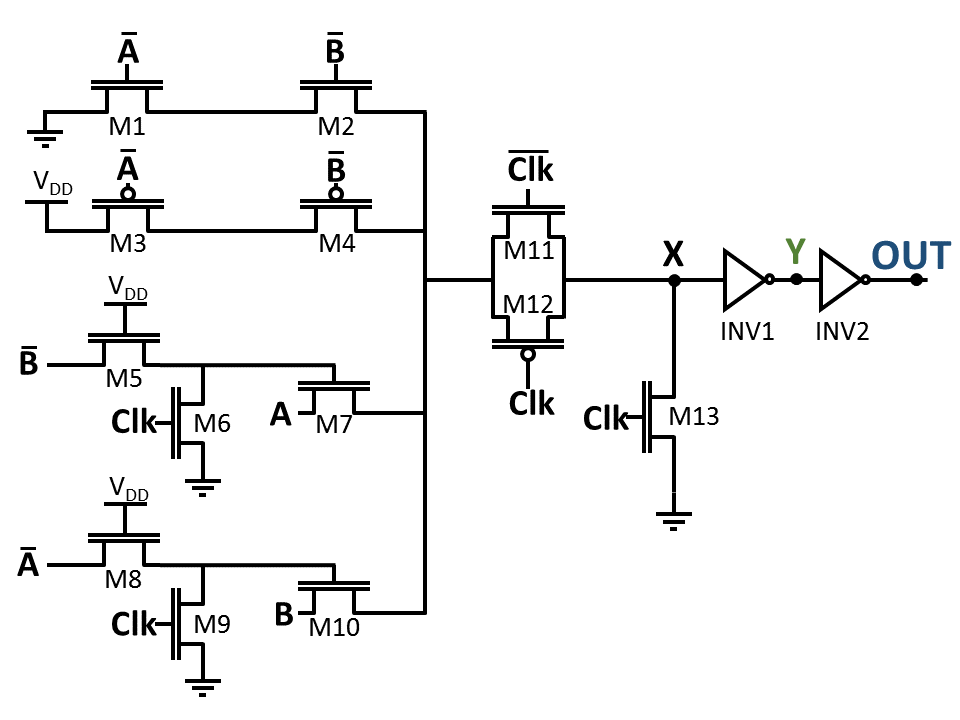}
 \caption{Proposed \abb{} schematic that functions as an AND for High-$V_{Th}$ and as an OR operation for Low-$V_{Th}$.}
\label{fig:camo-gate}
\end{figure}

Figure~\ref{fig:camo-gate} shows the schematic of our implementation of a camouflaged AND/OR gate. It is a \textit{clocked} logic family (similar to domino logic). Our implementation has two phases of execution, a predischarge (PRE) and an evaluate (EVAL) phase. In the EVAL phase the clock CLK signal is logic 0, and it is 1 during the PRE phase.

The logic function is determined by the activation of one of the four cases (branches), each dealing with a unique possible input combination. The top two branches of the layout correspond to inputs $A=B=0$ and $A=B=1$. We observe that the output is the same regardless of the function of the cell. For these cases transistors $M_1$ and $M_2$ are on while transistors $M_3$ and $M_4$ are off, or vice versa. The third branch corresponds to $A=1$ and $B=0$. Here we notice that the output changes depending on whether the desired functionality is an OR or an AND cell. In order to be able to obtain different behaviour, we connect NMOS transistors $M_5$ and $M_7$ in the aforementioned cascaded layout. Since we have the same case for $A=1$ and $B=0$, we connect NMOS transistors $M_8$ and $M_{10}$ in the same manner for last branch. In summary, we use the cascade configuration when we want to be able to pull down to zero a specific input combination depending on the manufacturing process.

\begin{figure}
\centering
\begin{subfigure}{.49\columnwidth}
\includegraphics[width=1.8in, height=1in]{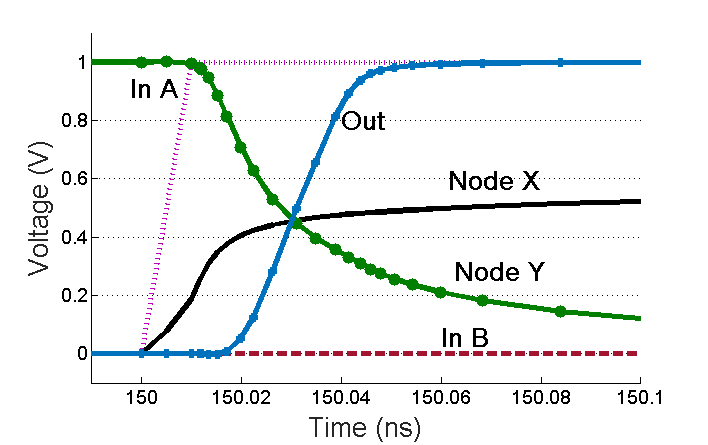}
\end{subfigure}%
~
\begin{subfigure}{.49\columnwidth}
\includegraphics[width=1.8in, height=1in]{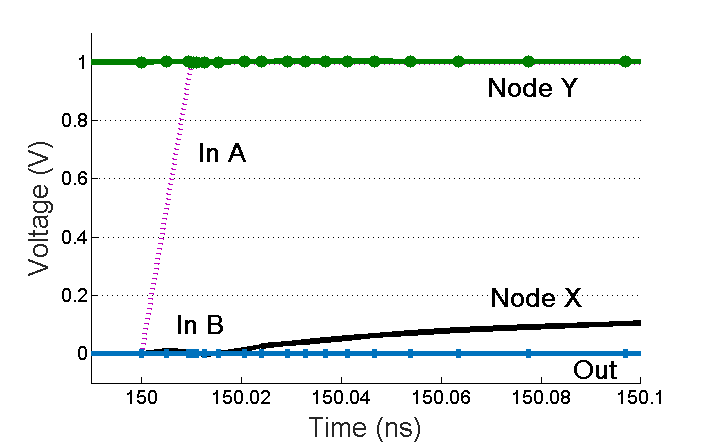}
\end{subfigure}%
\caption{(Left) Low $V_{Th}$ voltage response of internal nodes of \abb{} (Fig.~\ref{fig:camo-gate}) to inputs $A = 0 \to 1$ and $B=0$. Node X rises past 0.45V, causing node Y to fall, which in turn forces the output to rise. This circuit behaves like an OR gate. (Right) High $V_{Th}$ voltage response of internal nodes of \abb{} (Fig.~\ref{fig:camo-gate}) to inputs $A = 0 \to 1$ and $B=0$. Node X does not pass 0.11V and does not affect node Y. The output remains the same. This circuit behaves like an AND gate.}
\label{fig:response}
\end{figure}

This different behaviour can be visualized in Figure~\ref{fig:response}. 
In the left, the voltage at Node X rises to $0.45~V$, while in the right the voltage rises to $0.11~V$. We determine that the first voltage is a logic 1 and the lower voltage is a logic 0 by setting the mid-point voltage of the succeeding inverter, INV1, between $0.11~V$ and $0.45~V$. In this manner we can observe an accentuated difference at Node Y.  In Figure~\ref{fig:response} (Left), the voltage at Node Y falls below $0.2~V$, while in Figure~\ref{fig:response} (Right) the voltage remains close to $V_{DD}$. The final inverter, INV2, also skewed, provides clean logic levels at the output. 

\subsubsection{Clocked Logic}

So far we have assumed, implicitly, that Node X is initialized to $0 V$, and either stays at $0 V$ or transitions to $V_{DD}$. 
However, if  Node X starts at $V_{DD}$ then the output for the 
$A=1$ and $B=0$ case, as well as for the $A=0$ and $B=1$, will stay at $V_{DD}$ for the High-$V_{Th}$ process. For correct operation as an AND gate, the output is expected to fall to $0 V$. In order to deal with this complication we add NMOS transistors that pre-discharge the affected nodes. This technique is similar to the one used in dynamic logic cells. When $Clk = 1$, Node X is initialized to $0 V$.  We also need to pre-discharge the outputs of transistors $M_5$ and $M_8$ to eliminate accumulated charge at these nodes over multiple clock cycles. Our schematic is composed of the clocking transistors $M_6, M_9, M_{11}, M_{12},$ and $ M_{13}$. They are necessary to discharge the nodes periodically. Lastly, \abb{} operate as expected when the inputs transition from $0 \rightarrow 1$ in the EVAL phase, which is like conventional dynamic logic. On the other hand, if the  input transitions from $1 \rightarrow 0$ during the EVAL phase, the output for the High-$V_{Th}$ case may be incorrect. Consequently the proposed \abb{} are compatible with domino logic implementations, but cannot be used with static CMOS logic.

\subsection{Integration With Domino Logic}
Domino logic~\cite{rabaey-03} is a clocked logic design family in which each dynamic logic gate is followed by an inverting gate, for example, a static CMOS inverter. Domino logic has been used in commercial processor designs where speed is of the essence~\cite{stackhouse} and work on improving performance of dynamic logic \cite{Jeyasingh11} is ongoing.
Domino logic gates implement non-complemented Boolean functions like AND and OR instead of complemented Boolean functions (such as NAND and NOR) like static CMOS. Therefore, once a $0 \rightarrow 1$ transition occurs at the input, this will result in a $0 \rightarrow 1$ transition at the output during the EVAL stage. A domino logic implementation netlist can be camouflaged using \abb{} by simply \emph{replacing} conventional domino AND and domino OR 
gates with the proposed camouflaged AND/OR gate (Figure~\ref{fig:mini-bntlist}).

In Figure~\ref{fig:mini-bntlist}, the designer has chosen to camouflage two gates using \abb{}. By using reverse engineering techniques the attacker is unable to identify whether the \abb{} are ANDs or ORs, therefore the camouflaged netlist could implement one of four different Boolean logic functions: (i) $f =  ABC + \overline{B}C$; (ii) $f =  C$; (iii) $f =  AC\overline{B}$; and (iv) $f = 0$. The set of possible functions that a netlist with \abb{} could implement depends on the set of gates that are chosen. We note, however, that the question of how many and which gates to camouflage is out of the scope of our work and is not addressed here.

\subsection{Universal Logic Family}

Like Domino logic, \abb{} implement non-inverting logic gates. Therefore, a reasonable question to ask is --- can any Boolean logic function 
be implemented using \abb{}? The answer is \emph{yes}. This is 
done by ensuring that the 
primary inputs to the circuit are available in both true and complemented forms. 
Now, any Boolean logic function can be expressed in \emph{sum-of-product} form using only true literals if the 
true and complemented 
form of every input is considered to be a separate variable. In other words, the function can be implemented using AND and OR gates if both 
true and complemented of the primary inputs are available.

\begin{figure}[tpb]
\centering
\includegraphics[width=3.2in, height=2.2in]{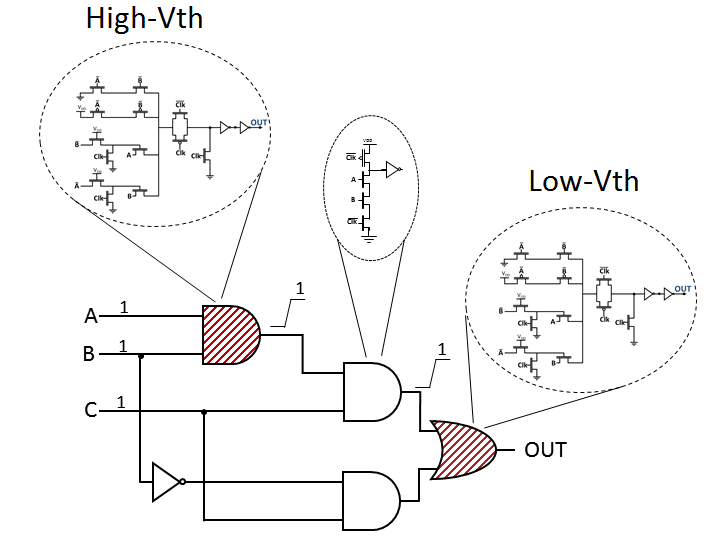}
 \caption{The camouflaged cells integrate with the rest of the design with their appropriate function, yet through visual inspection of the layout the individual function of each cell cannot be determined because they share the same exact structure.}
\label{fig:mini-bntlist}
\end{figure}

Observe that although the complemented primary inputs can 
result in  $1 \rightarrow 0$ transition, these transitions can be guaranteed to occur during the precharge phase since they occur immediately after the clock edge. For instance, in the example netlist in Figure~\ref{fig:mini-bntlist}, observe that a $1 \rightarrow 0$ transitions only occurs at the output of the static CMOS inverter driven by primary input $B$. 
Provided that the propagation delay of the static CMOS inverter is less than the half period of the clock, these transitions will occur in the PRE phase and will not cause problems. All other transitions in the netlist only occur in the EVAL phase and are always $0 \rightarrow 1$ transitions.

\subsection{Robustness to Process Variation}
Since our \abb{} depend on the threshold voltage, we must have careful consideration to the robustness to manufacturing process variations. Variations in doping concentration, which directly affect the operating threshold voltage, should not cause incorrect behaviour of the cell. In particular, even if the high-$V_{Th}$ ($V_{Th,H}$) and low-$V_{Th}$ ($V_{Th,L}$) implementations vary around their nominal values, \abb{} should continue to operate as AND and OR gates, respectively. We should specially consider the \textit{slow corner} for $V_{Th,L}$ and the \textit{fast corner} for $V_{Th,H}$. Assuming that $\Delta V_{Th}$ is the maximum difference (positive or negative) in threshold voltage, that ($V_{Th,L} + \Delta V_{Th}$) characterizes the slow corner for low-$V_{Th}$, and that ($V_{Th,H} - \Delta V_{Th}$) characterizes the fast corner for High-$V_{Th}$, for correct operation equation \ref{eq:condition} is a condition that must be satisfied. \vspace{-3ex}

\begin{equation}
(V_{Th,L} + \Delta V_{Th}) < (V_{Th,H} - \Delta V_{Th})
\label{eq:condition}
\end{equation}

To satisfy \ref{eq:condition}, the mid-point voltage of INV1 must be set between $V_{Th,L} + \Delta V_{Th}$ and $V_{Th,H} - \Delta V_{Th}$. We can achieve this by skewing INV1 so that the cut-off threshold voltage is suitable for detecting the logic 0 or 1 value that is reached by Node X in Fig.~\ref{fig:camo-gate}. 

In Section~\ref{sec:exp} we establish a range of skew values for which the cell works correctly, including the aforementioned corner cases.

\section{Experimental Results}\label{sec:exp}

We have used 22 $nm$ technology libraries from the ASU Process Technology Model (PTM) repository~\cite{ptm} for implementing our \abb{}. The 22 $nm$ technology has customizable predictive model files for both high performance (low-$V_{Th}$) and low power (high-$V_{Th}$) transistors. We used the HSPICE circuit simulation tool for all simulations. 
The nominal supply voltage for this technology is $V_{DD} = 1~V$. The $3\sigma$ variability in doping concentration 
is assumed to be $15\%$ of the nominal doping concentration 
for both the low power and high performance libraries. 

First, we characterize the delay, power consumption and robustness of 
our proposed \abb{}. Then we present a case study by integrating \abb{} into a large benchmark netlist and analyze delay and power overheads as a function of the number of camouflaged gates.

\begin{figure}[htpb]
\centering
\includegraphics[width=3.2in,height=1.8in]{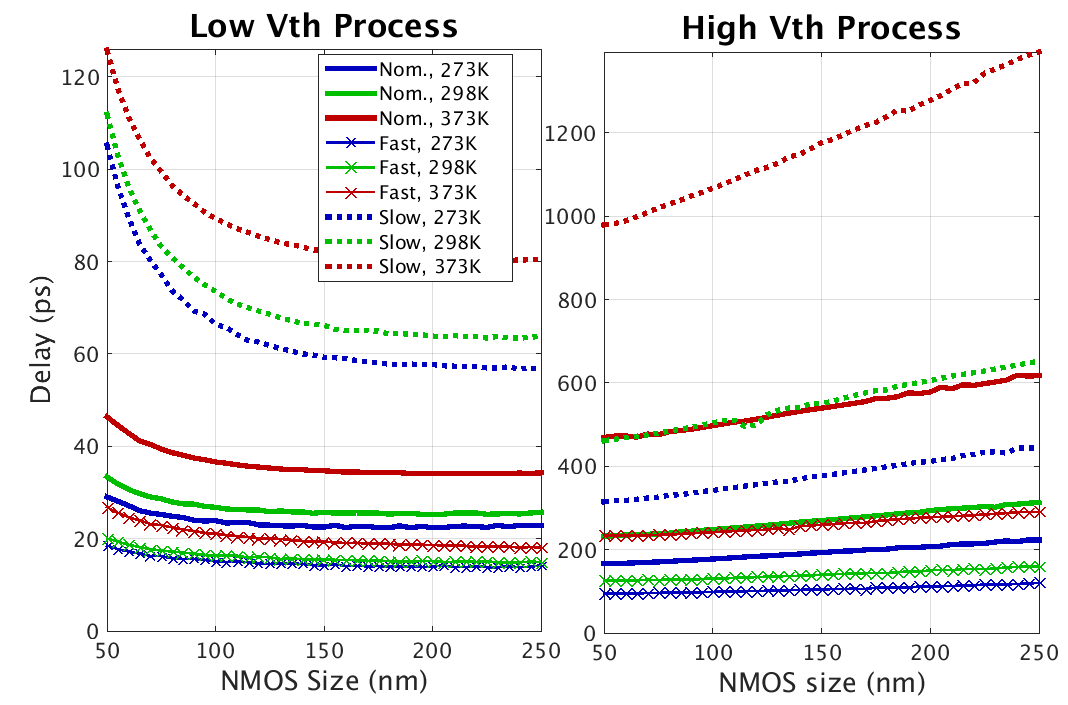}
 \caption{Delay as a function of size of NMOS width of INV1 for all nine PT (process and temperature corners) of both low-$V_{Th}$ and high-$V_{Th}$ \abb{}.}
\label{fig:pvt-skew}
\end{figure}

\subsection{Robustness  of \abb{}}
Our first goal is to size transistors so as to guarantee robustness against PVT variations. To do so, we considered $\pm 15\%$ process variations in threshold voltage \footnote{This assumption is justified by results in \cite{Cheng09} that indicate a
$3\sigma$ of variation in threshold voltage to be 90 mV at 22 nm.}, 
temperature ranging from $273K$ to $373K$ and a $V_{dd}$ range from  $0.85V$ to $1.15V$. 
Figure~\ref{fig:pvt-skew} 
shows the delay of the \abb{} on the y-axis versus size of the NMOS transistor in INV1 (for a fixed PMOS size) on the x-axis for different PVT corners.
\color{black}
We find that the for NMOS sizes ranging between 50 $nm$ and 250 $nm$, 
the cell works correctly over \emph{all PVT corners} and for both the the low-$V_{th}$ and high-$V_{th}$ cases. We therefore picked an NMOS INV1 size of 80 $nm$.

\subsection{Power and Delay Characterization of \abb{}}
In Table~\ref{delay_power_table} we present the propagation delay and power consumption of the  proposed \abb{} as well as conventional domino logic AND and OR gates. \abb{} with OR function (implemented with low-$V_{Th}$ transistors) have a delay $5.16\times$ greater than a low-$V_{Th}$ domino OR. Also, when a \abb{} functions as an AND gate (implementation with high-$V_{Th}$ transistors), it presents a delay that is $4.08\times$ greater than a high-$V_{Th}$ domino AND. From a power perspective, we note that the overheads for the \abb{} are $1.39\times$ and $1.04\times$ the corresponding values for the domino AND and OR gates, respectively.

\setlength\tabcolsep{1pt}
\begin{table}[ht]  
\small
\centering
\caption{Delay and Power Characteristics of \abb{} (TDCC)}
    \label{delay_power_table}

  \begin{tabular}{lccccccccc}
    \multirow{2}{*}{} &&&&
      \multicolumn{2}{c}{Static CMOS} & \multicolumn{2}{c}{Domino} & \multicolumn{2}{c}{TDCC}\\
    \hline
     & & & & {AND} & {OR} & {AND} & {OR} & {AND} & {OR} \\
      \hline \hline
      \multirow{2}{*}{ \parbox{.3cm} {\centering {Delay $(ps)$}}} 
      & \multicolumn{3}{c}{Low $V_{th}$}  & {7.68}  & {11.26} & {9.93} & {4.98} & {--} & {25.7}\\
      & \multicolumn{3}{c}{High $V_{th}$} & {34.03} & {50.47} & {66.00} &{23.63} & {269.8} & {--}\\
      \hline
       \multirow{9}{*}{ \parbox{.3cm} {\begin{turn}{90}{Power $(\mu W)$}\end{turn}}} & & {$A$}&{$B$}        &  &  &  & \\
        \cline{3-4}                     
       & \multirow{4}{*}{\parbox{0.4cm} {\centering Low $V_{th}$} }
                               & {$0$}&{$0$}        & {0.10} & {0.12} & {0.10} & {0.29} & {-} & {2.49}\\
                             & & {$0$}&{$0\to1$}    & {0.57} & {7.32} & {0.11} & {9.17} & {-} & {8.64}\\
                             & & {$0\to1$}&{$0$}    & {0.13} & {5.92} & {0.37} & {9.17} & {-} & {8.07}\\
                             & & {$0\to1$}&{$0\to1$} &{7.93} & {6.61} & {5.81} & {8.94} & {-} & {9.41}\\
                             \cline{2-10}
                            & \multirow{4}{*}{\parbox{0.4cm} {\centering High $V_{th}$} }
                               & {$0$}&{$0$}        & {0.00} & {0.00} & {0.02} & {0.01} & {0.13} & {-}\\
                             & & {$0$}&{$0\to1$}    & {0.96} & {1.55} & {0.01} & {3.36} & {0.37} & {-}\\
                             & & {$0\to1$}&{$0$}    & {0.84} & {1.48} & {0.13} & {3.36} & {0.38} & {-}\\
                             & & {$0\to1$}&{$0\to1$} &{1.29} & {1.49} & {1.66} & {3.78} & {1.65} & {-}\\
    \hline
  \end{tabular}
  
\end{table}

\subsection{Results of Netlist Integration of \abb{}}

To provide a better understanding of the impact of the delay and power overheads caused by \abb{} we have integrated our camouflaged logic circuits into a benchmark netlist. The logic circuits are first implemented using domino logic after which a randomly chosen subset of gates is camouflaged. We have used random selection~\cite{epic} in this work, yet other selection algorithms may also be used~\cite{RSS+13}.

We have implemented an automated toolflow using the Berkeley SIS synthesis framework~\cite{sentovich1992sis}. It's input is combinational logic described in .pla format and outputs an implementation that can be implemented using domino logic (similar to Figure~\ref{fig:mini-bntlist}). To do this, we replace the complemented form of each variable with a new variable and input the new .pla file to SIS. The netlist that is used has non-complemented gates, and we further restrict our current implementation to the use of just AND and OR gates.

For our netlist integration of \abb{} we have used the benchmark apex5 from the MCNC benchmark suite. This benchmark has 234 inputs, 37 outputs, and 2382 product terms. The results are normalized to the uncamouflaged domino logic baseline implementation. For consistency, all AND gates were implemented using high-$V_{Th}$ and all OR gates were implemented using low-$V_{Th}$, even though other selection criteria may be used. As expected, Table~\ref{delay_power_table} presents results obtained from simulations. With 40\% of the gates camouflaged the delay overhead is around 72.7\% and power overhead is near 31\%. These results may vary depending on which set of gates are chosen to be camouflaged and which non camouflaged gates will be fabricated with which process (high performance or low power). It is necessary to take into consideration the delay and power overheads as well as the critical path.

\begin{table}[h]
\caption{Delay and Power Overhead of \abb{}}
\label{delay_power_table}
\begin{tabularx}{\columnwidth}{c *{4}{>{\centering\arraybackslash}X}}   
 \hline
  & \multicolumn{4}{c}{Percent of Camouflaged Cells in Netlist}\\
 \cline{2-5}     
  & 10\% & 20\% & 30\% & 40\%\\
 \hline
 Delay & 1.3613  & 1.5153  & 1.6305 & 1.7270\\
 \hline
 Power & 1.0770  & 1.1551  & 1.2327 & 1.3105\\
\hline

\end{tabularx}

\end{table}

\section{Possible Attacks}
Like other camouflaging solutions, there are potential approaches 
that attackers could use to determine the identities of camouflaged 
gates. These include measuring the etch rate during chemical deprocessing, which could be used to determine which which transistor is more heavily doped.
In addition, an attacker might use the difference in power and delay 
characteristics of camouflaged gates to determine their identities. 
However, all camouflaging techniques are susceptible to this attack to some degree. Finally, attackers might develop 
more accurate techniques to measure the doping concentration of 
transistor channels, but this represents that natural cat-and-mouse game between attackers and defenders.

\section{Conclusion}\label{sec:conc}

This work proposes a new threshold dependent camouflaging technique that protects digital ICs against delayering and optical imaging based IC reverse engineering. Existing solutions consist of camouflaging Boolean functionality of a cell via the placement of true and dummy contacts, therefore they posses a vulnerability since they can be identified depending on the difference of the contacts. However, our proposed \abb{} only differ in operating threshold voltage and have the same physical structure regardless of their Boolean functionality. The threshold voltage of the cell depends on the inherent doping concentration of the transistor which cannot be determined by using traditional reverse engineering optical imaging techniques. Although other means for determining doping concentration exist, they have limited spatial resolution or accuracy. 

We have presented an analysis of delay and power consumption overhead for our proposed \abb{}. This IC camouflaging alternative is based on a design that leverages the threshold-voltage dependent level shifting property cascaded pass gates, and is shown to be robust against process variations incurred during manufacturing. \abb{} can be seamlessly incorporated into a netlist with domino logic gates. Experimental results indicate that camouflaging up to $40\%$ of gates of the apex5 benchmark induced delay and power overheads near than $70\%$ and $30\%$, respectively.

\section{Future Work}

Consequently, our proposed \abb{} increase the effort necessary for attackers who wish to reverse engineer the netlist of a digital IC. 
Future work will focus on designing a more comprehensive library of \abb{}, further optimizing the delay, as well as power and robustness of \abb{}. Understanding vulnerabilities that may come from the analysis of side channels as well as defending against advanced attacks is crucial.

\color{black}

\section*{Acknowledgment}

The authors acknowledge funding support from the NSF CNS Award \#1527072, NSF CNS Award \#1553419, and the Semiconductor Research Corporation.



%

\bibliographystyle{ieeetr}
\bibliography{allrefs,ndssrefs}

\end{document}